\documentclass[sigconf]{acmart}

\usepackage[all]{nowidow}
\usepackage{balance}
\usepackage{anyfontsize}

\usepackage{array}
\usepackage{multirow}
\usepackage{booktabs}
\usepackage{tabularx}
\usepackage{makecell}
\usepackage{adjustbox}
\usepackage{graphicx}
\usepackage{varwidth}
\usepackage{colortbl}

\usepackage{xcolor}
\definecolor{darkgreen}{rgb}{0.09, 0.45, 0.27}
\definecolor{darkbrown}{rgb}{0.59, 0.29, 0.0}
\definecolor{commentGray}{RGB}{120,120,120}
\definecolor{light-gray}{gray}{0.9}
\definecolor{amplcol}{HTML}{DDEBF7} % light blue
\definecolor{pycol}{HTML}{E2F0D9}   % light green
\definecolor{javagreen}{rgb}{0.25,0.5,0.35}

\usepackage{algorithm}
\usepackage{algorithmicx}
\usepackage{algpseudocode}

\renewcommand{\algorithmiccomment}[1]{\bgroup\color{commentGray}{//#1}\egroup}

\usepackage{listings}
\lstdefinestyle{Alg}{
  basicstyle=\ttfamily\footnotesize,
  breaklines=true,
  tabsize=2,
  mathescape,
  numbers=left,
  xleftmargin=2.5em,
  xrightmargin=0.5em,
  frame=tb,
  framexleftmargin=2em,
  emph={Algorithm,Input,Output,for,each,do,if,else,Function,while,let,be,repeat,until,return,times,and,or,break,in,then,},
  emphstyle={\textbf},
  escapechar=?,
  morecomment=[l][\color{javagreen}]{//},
  columns=flexible,
}

\usepackage{framed}
\usepackage{tikz}
\usepackage[framemethod=TikZ]{mdframed}
\mdfsetup{skipabove=5pt,skipbelow=3pt}
\mdfdefinestyle{RQFrame}{%
	linecolor=black,
	outerlinewidth=0.15pt,
	roundcorner=3pt,
	innertopmargin=2pt,
	innerbottommargin=2pt,
	innerrightmargin=4pt,
	innerleftmargin=4pt,
	backgroundcolor=light-gray}

\usepackage[most]{tcolorbox}
\tcbuselibrary{skins,breakable,listings}
\newtcblisting[blend into=figures]{PromptBox}[2][]{%
  enhanced, breakable,
  listing engine=listings, listing only,
  colback=gray!3, colframe=black!20, boxrule=0.4pt, arc=1.5mm,
  title={#2},
  listing options={
    basicstyle=\ttfamily\small,
    breaklines=true, columns=fullflexible,
    showstringspaces=false
  },
  #1
}

\lstdefinestyle{amplCompact}{
  basicstyle=\fontsize{5}{5}\selectfont\ttfamily
}

\usepackage{amsmath}
\usepackage{esvect}
\usepackage{pifont}

\newcommand{\uptriangle}{\textcolor{green!60!black}{$\blacktriangle$}}
\newcommand{\downtriangle}{\textcolor{red}{$\blacktriangledown$}}
\newcommand{\numup}[1]{#1\,\uptriangle}
\newcommand{\numdown}[1]{#1\,\downtriangle}
 
\newcommand{\cmark}{\ding{51}}%
\newcommand{\xmark}{\ding{55}}%

\newcolumntype{Y}{>{\centering\arraybackslash}X}

\newcommand{\approach}{EXEOS}
\newcommand{\llm}{\textsf{LLM}}

\newcounter{commentnumber}

\copyrightyear{2026}
\acmYear{2026}
\setcopyright{cc}
\setcctype{by}
\acmConference[ICSE-SEIP '26]{2026 IEEE/ACM 48th International Conference on Software Engineering}{April 12--18, 2026}{Rio de Janeiro, Brazil}
\acmBooktitle{2026 IEEE/ACM 48th International Conference on Software Engineering (ICSE-SEIP '26), April 12--18, 2026, Rio de Janeiro, Brazil}
\acmDOI{10.1145/3786583.3786879}
\acmISBN{979-8-4007-2426-8/2026/04}

\begin{document}

\title[DSL or Code? Evaluating the Quality of LLM-Generated Algebraic Specifications]{DSL or Code? Evaluating the Quality of LLM-Generated Algebraic Specifications: A Case Study in Optimization at Kinaxis}

\author{Negin Ayoughi}
\affiliation{%
  \institution{University of Ottawa \& Kinaxis}
  \city{Ottawa}
  \state{Ontario}
  \country{Canada}
}
\email{negin.ayoughi@uottawa.ca}

\author{David Dewar}
\affiliation{%
  \institution{Kinaxis}
  \city{Ottawa}
  \state{Ontario}
  \country{Canada}
}
\email{ddewar@kinaxis.com}

\author{Shiva Nejati}
\affiliation{%
  \institution{University of Ottawa}
  \city{Ottawa}
  \state{Ontario}
  \country{Canada}
}
\email{snejati@uottawa.ca}

\author{Mehrdad Sabetzadeh}
\affiliation{%
  \institution{University of Ottawa}
  \city{Ottawa}
  \state{Ontario}
  \country{Canada}
}
\email{m.sabetzadeh@uottawa.ca}

\begin{abstract}
Model-driven engineering (MDE) provides abstraction and analytical rigour, but industrial adoption in many domains has been limited by the cost of developing and maintaining models. Large language models (LLMs) help shift this cost balance by enabling the direct generation of models from natural-language (NL) descriptions. For domain-specific languages (DSLs), however, there is the risk that LLM-generated models may be less accurate than LLM-generated code in mainstream languages such as Python, due to the latter's dominance in LLM training corpora. We investigate this issue in the domain of mathematical optimization, focusing on AMPL -- a DSL with established industrial use. We introduce \approach, an LLM-based approach that derives AMPL models and Python code from NL problem descriptions and iteratively refines them using solver feedback. Using a public optimization dataset and real-world supply-chain cases from our industry partner, Kinaxis, we evaluate how generated AMPL models compare with Python code in terms of executability and correctness. An ablation study across two LLM families shows that AMPL is competitive with -- and sometimes better than -- Python, and that our design choices in \approach\ improve the quality of generated specifications.
\end{abstract}

\begin{CCSXML}
<ccs2012>
  <concept>
    <concept_id>10011007.10010940.10010971.10010980.10010984</concept_id>
    <concept_desc>Software and its engineering~Model-driven software engineering</concept_desc>
    <concept_significance>500</concept_significance>
  </concept>
  <concept>
    <concept_id>10011007.10011006.10011050.10011017</concept_id>
    <concept_desc>Software and its engineering~Domain specific languages</concept_desc>
    <concept_significance>300</concept_significance>
  </concept>
  <concept>
    <concept_id>10011007.10011006.10011060.10011690</concept_id>
    <concept_desc>Software and its engineering~Specification languages</concept_desc>
    <concept_significance>300</concept_significance>
  </concept>
  <concept>
    <concept_id>10010147.10010257.10010293</concept_id>
    <concept_desc>Computing methodologies~Machine learning approaches</concept_desc>
    <concept_significance>300</concept_significance>
  </concept>
  <concept>
    <concept_id>10010147.10010178.10010179</concept_id>
    <concept_desc>Computing methodologies~Natural language processing</concept_desc>
   <concept_significance>100</concept_significance>
</concept>
</ccs2012>
\end{CCSXML}

\ccsdesc[500]{Software and its engineering~Model-driven software engineering}
\ccsdesc[300]{Software and its engineering~Domain specific languages}
\ccsdesc[300]{Software and its engineering~Specification languages}
\ccsdesc[300]{Computing methodologies~Machine learning approaches}
\ccsdesc[100]{Computing methodologies~Natural language processing}

\keywords{Model-Driven Engineering (MDE), Large Language Models (LLMs), Automated Model Extraction, Domain-Specific Languages (DSLs), Mathematical Optimization.}

\maketitle

\section{Introduction}
\label{sec:intro}
Model-driven engineering (MDE) has long been recognized for the conceptual abstraction and analytical rigour it brings to the development of software-intensive systems. Yet, outside sectors with certification or strict assurance mandates, the economics of building and maintaining models have often limited broader industrial uptake; many organizations perceive the costs as outweighing the benefits~\cite{Hutchinson11,Whittle13,Pourali18}. Large language models (LLMs) are rapidly changing this cost calculus. By reducing the effort required to generate and modify structured artifacts from textual descriptions, LLMs enable domain experts to work primarily in natural language (NL) while still preserving the benefits of modelling,  without incurring the high associated costs. This flexibility has led to growing interest in LLM-assisted derivation of models from text, e.g., \cite{FerrariAA24, YangCCMV24,Chen25,khamsepour2025impact}.

A new trade-off has nonetheless emerged: when models are intended as computational or behavioural specifications -- particularly in domain-specific languages (DSLs) that are less represented in LLMs' pretraining corpora -- the quality of LLM-generated models may fall short of LLM-generated code in mainstream languages like Python or Java, where the desired computation or behaviour is expressed directly in the program. In practice, this can give rise to a tension: although models seem preferable for a variety of reasons, such as abstraction and understandability, conventional code may better capture developer intent with fewer defects, given the prevalence of popular programming languages in LLM training data. The question many teams face is therefore not one of ``models or code in principle'', but rather which approach produces fewer issues and greater speed for a given task and context.

Our work in this paper is informed by an industrial collaboration with Kinaxis, a global provider of supply-chain planning software. There, analysts frequently need to represent business logic and decision-policy requirements as mathematical optimization problem specifications. Being able to express these requirements in text and automatically generate specifications from them is advantageous, not only because domain specialists --  who are not always experts in writing formal specifications -- supply them, but also because text is the de-facto medium for communicating with external stakeholders, such as clients, which is often essential as requirements are being elaborated. Having written the requirements in text, analysts must then determine whether transitioning to higher-level models or conventional code will provide greater benefits.

The issue of representation for LLM-generated specifications is not unique to the industrial setting that motivates our work. Similar questions can arise in other domains, where one must decide whether to generate models in a specialized language, or instead produce code directly in a general-purpose programming language. Examples of such domains, among many others, include transformation~\cite{ATL08}, system configuration~\cite{Dolstra10,Puppet}, orchestration~\cite{Monsanto12},  testing~\cite{gherkin}, and rule checking~\cite{datalog}. Under delivery pressure, teams may prefer ``touching up'' LLM-generated code (e.g., in Python) rather than revising a higher-level LLM-generated model that requires more substantial fixes. In other words, the potentially stronger out-of-the-box performance of LLMs in generating code from text makes code a tempting shortcut, creating the risk of a shift away from models to code even though models are now less costly to build through LLM assistance.

The central focus of this paper is \emph{whether LLM-generated DSL models can achieve parity with LLM-generated code in the qualities that matter most to practitioners: fidelity to stated intent and a low defect rate.} To explore this question, and reflecting the industrial context of our collaborating partner, we focus on AMPL (A Mathematical Programming Language) — a high-level, algebraic modelling language designed for capturing and solving large-scale mathematical optimization problems, such as linear programming, mixed-integer programming, and nonlinear programming~\cite{ampl, fourer1990modeling}. AMPL has strong traction at our industry partner, thus providing a concrete setting for studying the trade-offs between DSL model generation and code generation using LLMs.

\textbf{\emph{Contributions.}} Our main contributions are as follows:

\textbf{(1)} We introduce \approach\ (\textbf{EX}traction and \textbf{E}rror-guided refinement of \textbf{O}ptimization \textbf{S}pecifications), an LLM-based approach for deriving formal specifications of optimization problems from NL problem descriptions. \approach\ first structures the given description by identifying the main components of an optimization problem, then generates candidate specifications (in either AMPL or Python) and iteratively refines them using error diagnostics from a solver. A key and novel feature of \approach\ is its explicit handling of data, which is indispensable in industrial applications where parameter values are typically maintained in external databases rather than embedded in problem statements. This separation not only reflects practical realities but also focuses the role of LLMs on model construction --~where automation is actually needed~-- thereby reducing noise from data extraction and enabling more accurate automation.

\textbf{(2)} We empirically evaluate \approach, comparing AMPL models against Python code generated from identical inputs. Our evaluation is based on two datasets: a public corpus of optimization problems~\cite{Bertsimas24,williams2013model,Nace20} and a set of real-world cases from Kinaxis in supply-chain planning. For each problem instance, we generate both an AMPL model and Python code, considering all combinations of the structuring step (present or absent) and iterative refinement (enabled or disabled) in \approach. This factorial design amounts to an ablation study, allowing us to isolate the effects of the target language (AMPL vs. Python), the structuring step, and repair. To ensure robustness, we experiment with two LLM families -- GPT and Gemini -- spanning both reasoning and instruction-following LLMs, and repeat each run several times, yielding over \emph{eleven thousand specifications}. We assess executability by whether the specifications compile without errors, and correctness by comparing solutions to the ground truth using both exact matches and relative error.

\textbf{\emph{Findings.}} The main findings from our evaluation are as follows: First, introducing a structuring step prior to generating a formal specification consistently improves quality, reducing compilation errors and yielding solutions more closely aligned with the intended optimization objectives. Second, iterative refinement further increases executability rates by enabling the automatic correction of specifications that initially failed at compile time or runtime. Third, generating Python code does \emph{not} provide a systematic advantage over generating AMPL models. Specifically, across both datasets and LLM families, LLM-generated Python code shows no statistically significant improvement in correctness over AMPL. When \approach\ structures the NL problem description prior to specification generation and iteratively refines the generated specification, Python provides no executability benefit either: AMPL models execute more successfully on the public dataset and perform on par with Python on the industry cases. Indeed, AMPL models produced through the structure-generate-refine process generally outperform Python, especially when reasoning LLMs are used. This suggests that DSL model generation remains competitive, and in many cases even advantageous, when performed by sufficiently capable LLMs. Finally, our approach outperforms a baseline for automated optimization code generation without a separate data handling step as common in the literature~\cite{AhmadiTeshniziG24}, achieving higher executability and correctness in AMPL models.

\textbf{\emph{Replication Package.}} All code, evaluation scripts, and experimental data for our public dataset are available online~\cite{neayoughi2025optigen}.

\textbf{\emph{Terminology.}} Throughout the paper, we use ``specification'' to refer to either an AMPL model or a Python program; ``model'' is reserved for non-code (AMPL) specifications. Large language models are referred to only as ``LLM(s)''.

\section{Motivation}
\label{sec:motivation}
Figure~\ref{fig:ProductionProblem} presents a production-planning problem in which resources must be allocated to manufacture products, subject to inventory and budget constraints, with the objectives of maximizing revenue and minimizing inventory costs (the colour coding is explained in Section~\ref{sec:approach}). A supply-chain analyst aiming to solve this problem typically has access to the underlying data -- such as inventory costs, purchase prices, and the type and quantity of resources required for each product -- and the business knowledge needed to analyze it, but often lacks the expertise, or finds it too time-consuming, to formally develop a specification that can automatically solve such problems.

\begin{figure}[t]
\centering
\Description{Text box that presents a production planning optimization problem statement with color-coded elements for objectives, parameters, decision variables, and constraints.}
\begin{tcolorbox}[colback=white, colframe=black, 
  boxrule=0.5pt, width=0.48\textwidth,
  left=3pt, right=3pt, top=3pt, bottom=2pt, boxsep=0pt]
\footnotesize
We consider a production-planning problem that involves manufacturing some \textcolor{darkgreen}{types of products}, each requiring specific \textcolor{darkgreen}{types of resources} for its production. Each resource has an \textcolor{darkgreen}{initial inventory}, and additional resource units may be purchased at \textcolor{darkgreen}{specified costs}, subject to a \textcolor{darkgreen}{total budget}. The effective availability of a resource is therefore the sum of its initial stock and purchased quantity, while any unused portion of this availability incurs an \textcolor{darkgreen}{inventory cost}.
The production of each product consumes \textcolor{darkgreen}{certain amounts of each resource, as specified in a requirement table}. Each product generates revenue through a given \textcolor{darkgreen}{selling price per unit.}
The objective is to determine \textcolor{red}{production quantities} and \textcolor{red}{the additional resource units} that \textcolor{magenta}{maximize total sales revenue} while \textcolor{magenta}{minimizing the holding costs of unused resources}, \textcolor{darkbrown}{subject to resource requirements, inventory availability, and the budget constraint}.
\end{tcolorbox}
\vspace*{-1em}
\caption{An example of a natural-language description of an optimization problem. Text in \textcolor{magenta}{pink} denotes objectives, \textcolor{darkgreen}{green} denotes parameters, \textcolor{red}{red} denotes decision variables, 
and \textcolor{darkbrown}{brown} denotes constraints.}
\label{fig:ProductionProblem}
\vspace*{-1em}
\end{figure}

Given their ability to generate formal specifications from textual descriptions, LLMs are natural candidates for automatically formulating optimization problems such as the one shown in Figure~\ref{fig:ProductionProblem}. There are two main ways to approach this task with an LLM: translating the description into a general-purpose programming language such as Python, or into a domain-specific optimization modelling language such as AMPL, which we briefly introduce in Section~\ref{sec:background}. Figures~\ref{fig:python-spec} and~\ref{fig:ampl-0} illustrate LLM-generated specifications in Python and AMPL, respectively, for the problem in Figure~\ref{fig:ProductionProblem}.
To represent the data -- which the user must provide in addition to the problem statement in Figure~\ref{fig:ProductionProblem} -- we supply it in a dedicated data file in AMPL, consistent with its requirement to separate model and data. In Python, we embed the data directly in the code for brevity, although it could just as well be stored externally (e.g., in JSON).

% ===== Python =====
\begin{figure}
\centering
\Description{Two code listings for the same production planning problem: a Python implementation using Gurobi and an AMPL model with a separate AMPL data file.}
% --- Python listing style ---
\lstdefinelanguage{Python}{
  morekeywords={import,from,as,for,if,else,True,False,return,print,in,with,def,class,try,except,raise,continue,break,pass,not,and,or, addVars, addConstrs, quicksum, setObjectiveN, optimize, addConstr },
  sensitive=true,
  morecomment=[l]{\#},
  morestring=[b]",
  morestring=[b]'
}
\lstset{emph={addVars,addConstrs,addConstr,quicksum,setObjectiveN,optimize},
        emphstyle=\color{blue}}
\lstdefinestyle{pyCompact}{
  language=Python,
  basicstyle=\ttfamily\scriptsize\linespread{0.9}\selectfont,
  keywordstyle=\color{blue},
  commentstyle=\color{gray},
  showstringspaces=false,
  moredelim=**[is][\color{red!80!black}]{@}{@},
  columns=flexible,
  keepspaces=true,
  numbers=left,
  numberstyle=\tiny\color{gray},
  numbersep=6pt
}

\begin{tcolorbox}[colback=white, colframe=black, boxrule=0.4pt, arc=2pt, width=0.48\textwidth, height=10cm, valign=center]
\begin{lstlisting}[style=pyCompact, basicstyle=\fontsize{5}{5}\selectfont\ttfamily]
import gurobipy as gp
from gurobipy import GRB
# Data 
PRODUCTS   = ["A", "B"]
RESOURCES = ["R1", "R2", "R3"]
@price@ = {"A": 10, "B": 15}
@inventory@ = {"R1": 8, "R2": 10, "R3": 3}
@hold@      = {"R1": 10, "R2": 0,  "R3": 10}
@buyCost@   = {"R1": 1,  "R2": 1,  "R3": 1}

# Required units of resource r per unit of product p
@unit@ = {
    "R1": {"A": 1, "B": 1},
    "R2": {"A": 1, "B": 2},
    "R3": {"A": 0, "B": 1},
}

@budget@ = 10.0

# Model
m = gp.Model("production_with_budget")

# Decision variables
@x@ = m.addVars(PRODUCTS, name="x", lb=0)                   # production
@y@ = m.addVars(RESOURCES, name="buy", lb=0)                # purchases
@leftover@ = m.addVars(RESOURCES, name="leftover", lb=0)    # unused inventory

#  Inventory constraint
m.addConstrs(
    (@inventory@[r] + @y@[r] - gp.quicksum(@unit@[r][p] * @x@[p] for p in PRODUCTS)
     == @leftover@[r]) for r in RESOURCES )

# Budget constraint
m.addConstr(gp.quicksum(@buyCost@[r] * @y@[r] for r in RESOURCES) <= @budget@, name="budget")

# Set objectives to maximize
m.ModelSense = GRB.MAXIMIZE

# Revenue objective (positive weight)
m.setObjectiveN(gp.quicksum(@price@[p] * @x@[p] for p in PRODUCTS), index=0, 
priority=1, weight=1.0, name="Revenue" )

# Holding cost objective (negative weight)
m.setObjectiveN(gp.quicksum(@hold@[r] * @leftover@[r] for r in RESOURCES), index=1, 
priority=0, weight=-1.0, name="Neg_HoldCost")

m.optimize()
\end{lstlisting}
\end{tcolorbox}
\vspace*{-1 em}
\caption{Python-based formulation of the production-planning optimization problem from Figure~\ref{fig:ProductionProblem}.}
\label{fig:python-spec}
%\vspace*{-.5em}
%\end{figure}
\vspace*{1.5em}
%\begin{figure}[!t]
\centering
\lstdefinelanguage{AMPL}{
  morekeywords={set, param, var, maximize, minimize, subject, to, sum},
  sensitive=true,
  morecomment=[l]{\#},
}

\lstdefinestyle{amplCompact}{
  language=AMPL,
  basicstyle=\ttfamily\scriptsize\linespread{0.9}\selectfont,
  keywordstyle=\color{blue},
  commentstyle=\color{gray},
  showstringspaces=false,
  moredelim=**[is][\color{red!80!black}]{@}{@},
  columns=flexible,
  keepspaces=true,
  numbers=left,
  numberstyle=\tiny\color{gray},
  numbersep=6pt
}
\lstset{
  emph={price,unit,inventory,hold,buyCost,budget,
        x,y,leftover,Balance,Budget_Limit,Revenue,Hold_Cost},
  emphstyle=\color{red!80!black}
}
\noindent

% Left box
\begin{minipage}[t]{0.31\textwidth}
\footnotesize \textbf{(a) AMPL Model}

\begin{tcolorbox}[colback=white, colframe=black, boxrule=0.4pt,
                  arc=2pt, width=\linewidth, height=5.8cm, valign=center]

\begin{lstlisting}[style=amplCompact, basicstyle=\fontsize{5}{5}\selectfont\ttfamily]
# Sets
set PRODUCTS; 
set RESOURCES; 
# Parameters 
param price {PRODUCTS} >= 0; 
param unit {RESOURCES, PRODUCTS} >= 0; 
param inventory {RESOURCES} >= 0; 
param hold {RESOURCES} >= 0; 
param buyCost {RESOURCES} >= 0; 
param budget >= 0; 
# Decision variables 
var x {p in PRODUCTS} >= 0;  # production
var y {r in RESOURCES} >= 0; # purchases
var leftover {r in RESOURCES} >= 0; # unused inventory
#  Inventory constraint
subject to Balance {r in RESOURCES}:
  inventory[r] + y[r] -
    sum {p in PRODUCTS} unit[r,p] * x[p] = leftover[r];
# Budget constraint 
subject to Budget_Limit:
  sum {r in RESOURCES} buyCost[r] * y[r] <= budget;
# Objective:  maximize revenue
maximize Revenue:  
  sum {p in PRODUCTS} price[p] * x[p];
# Objective:  minimize holding cost
minimize Hold_Cost:
  sum {r in RESOURCES} hold[r] * leftover[r]; 
\end{lstlisting}
\end{tcolorbox}
\end{minipage}
\hfill
% Right box
\begin{minipage}[t]{0.16\textwidth}
\footnotesize \textbf{(b) AMPL Data}

\begin{tcolorbox}[colback=white, colframe=black, boxrule=0.4pt,
                  arc=2pt, width=\linewidth, height=5.8cm, valign=center]
\begin{lstlisting}[style=amplCompact, basicstyle=\fontsize{5}{5}\selectfont\ttfamily]
set PRODUCTS := A B;
set RESOURCES := R1 R2 R3;
param price :=
  A 10 
  B 15;
param inventory :=
  R1 8
  R2 10
  R3 3;
param hold :=
  R1 10
  R2 0
  R3 10;
param buyCost :=
  R1 1
  R2 1
  R3 1;
param unit :
         A     B :=
    R1   1     1
    R2   1     2
    R3   0     1;
param budget := 10;
\end{lstlisting}
\end{tcolorbox}
\end{minipage}
\vspace*{-0.5em}
\caption{AMPL-based formulation of the production-planning optimization problem from Figure~\ref{fig:ProductionProblem}.}
\Description{Two side-by-side code listings. The left listing shows an AMPL model for a production-planning problem with sets, parameters, decision variables, inventory balance and budget constraints, and objectives for revenue and holding cost. The right listing shows an AMPL data instance with products, resources, prices, inventories, unit requirements, costs, and a budget.}
\label{fig:ampl-0}
\vspace*{-1.5em}
\end{figure}

Both the Python program in Figure~\ref{fig:python-spec} and the AMPL model in Figure~\ref{fig:ampl-0} are executable specifications of the optimization problem in Figure~\ref{fig:ProductionProblem}. The Python program is solver-specific, targeting the API of the Gurobi solver~\cite{gurobi}, whereas the AMPL model is solver-independent and can run with any compatible solver.
For domain experts, the AMPL model offers clear advantages: in addition to being solver-independent, it is more compact and generally more natural than Python code. The difference is most evident in how objectives are expressed. In the Python program, lines 36–47 encode the revenue and holding-cost objectives; this requires knowledge of the Gurobi API and some programming workarounds. In particular, because Gurobi enforces a single global optimization sense (either maximization or minimization), all objectives must conform to that requirement. Thus, when the solver is set to maximize, the holding-cost objective (which should naturally be minimized) must be reformulated with a negative weight. Although valid, this approach makes the Python code harder to interpret or manually adjust. By contrast, the AMPL model on lines 22–27 expresses the objectives in a form much closer to the original problem description.

In our example, if the user does not need to concern themselves with the effort to develop either the Python code or the AMPL model, the benefits of using a DSL such as AMPL would likely outweigh those of a general-purpose programming language like Python, because readability, solver independence, and alignment with the problem description reduce more long-term effort than general-purpose flexibility. However, this cost–benefit balance becomes less straightforward when LLMs generate candidate specifications. Specifically, because LLMs are trained on vast amounts of code in mainstream programming languages, including Python, there is a risk they perform better when generating Python implementations than specifications in a specialized DSL such as AMPL, which is far less represented in LLM training corpora. In such cases, one may face a practical trade-off: correcting a Python implementation already close to functional, or revising an AMPL model that, while more desirable if accurate, requires considerable effort to repair. Under time and budget pressures, this imbalance could cause practitioners to bypass models and opt for code instead.

Our goal is to examine this potential tension in an industrial setting, where specifications involve mathematical optimization, by having LLMs transform them into a formal representation -- written either in a general-purpose programming language (Python with optimization libraries) or in a DSL (AMPL). Our results show that, with careful design of the specification-derivation approach, automatically derived DSL specifications can achieve quality comparable to -- and in some cases exceeding -- that of code. In our study context, this would shift the balance in favour of modelling, even in the presence of LLMs.

\section{The AMPL Language}\label{sec:background}
Mathematical optimization problems (optimization problems for short hereafter) formally model decision-making tasks, enabling one to determine the optimal values of related variables in order to achieve specific objectives. 
A Mathematical Programming Language (AMPL)~\cite{ampl, fourer1990modeling} is a high-level modelling language designed for formulating and solving optimization problems. AMPL enables users to express optimization problems in a declarative, algebraic notation that closely resembles standard mathematical formulations. AMPL separates problem specification from data, allowing the same formulation to be applied to different datasets by updating only the data. AMPL also supports a broad range of solvers, enabling problems to be solved by different ones without altering their formulation. These flexible features have made AMPL suitable for use in both industry and academic research~\cite{ampl, fourer1990modeling}.

For example, Figure~\ref{fig:ampl-0} illustrates an AMPL-based representation of the problem  in Figure~\ref{fig:ProductionProblem}: Figure~\ref{fig:ampl-0}(a) shows the AMPL model, and Figure~\ref{fig:ampl-0}(b) provides the data with the concrete parameter values. The parameter names in the data file match those declared in the formulation in Figure~\ref{fig:ampl-0}(a). In Figure~\ref{fig:ampl-0}(a), the model declares parameters using \textcolor{blue}{\textsf{param}} on lines 5-10 and decision variables using \textcolor{blue}{\textsf{var}} on lines 12-14. The inventory and budget constraints are defined under \textcolor{blue}{\textsf{subject to}} on lines 16–21. The two objectives -- maximize revenue and minimize inventory costs -- are specified with \textcolor{blue}{\textsf{maximize}} and \textcolor{blue}{\textsf{minimize}} on lines 23–27.
\section{Our Approach}
\label{sec:approach}
\begin{figure} [t]
	\centering
    \Description{Diagram of the EXEOS pipeline with four steps: structuring and metadata extraction, data transformation, specification generation or refinement, and solver execution.}
    \hspace*{-.5em}\includegraphics[width=0.911\linewidth]{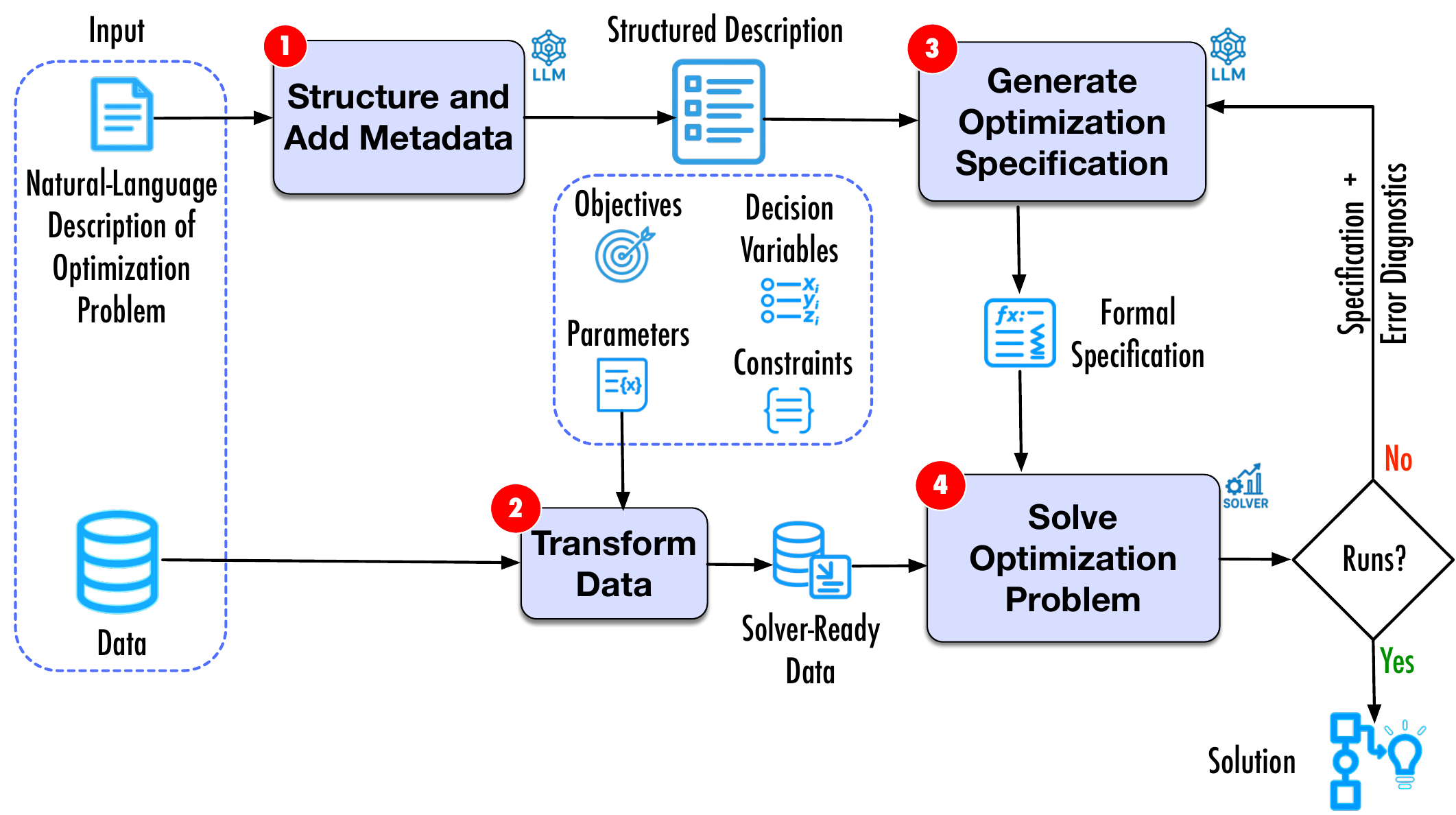}
        \vspace*{-1em}
        \caption{\approach\ --  our approach for transforming NL descriptions of optimization problems into formal specifications.}\label{fig:approach}
        \vspace*{-1em}
\end{figure}

Figure~\ref{fig:approach} outlines our approach, \approach, for translating NL descriptions of optimization problems into formal specifications that can be solved by existing solvers. The target specification language is configurable and can be either AMPL or Python.

\approach\ takes two inputs: (1)~an NL description of the optimization problem to be solved, and (2)~the underlying data, such as tables retrieved from a database. Based on these inputs, \approach\ produces a solution to the given optimization problem.

The approach has four steps: Step~1 structures the NL problem description and augments it with metadata. Step~2 processes the input data into a solver-ready data file. Step~3 generates a formal optimization specification, or regenerates it based on feedback from Step~4 if the solver fails due to compilation or runtime errors. In Step~4, the data file from Step~2 and the specification from Step~3 are used to compute a solution. Steps~1~and~3 require an LLM. Although different LLMs could be used, we employ a single instance for efficiency, given the sequential nature of the steps. We refer to this instance as \llm\ throughout the paper. The full list of prompts for \approach\ is available online~\cite{optigen-supp-a}. We next detail the steps of \approach.

\textbf{Step~1. Structure and Add Metadata.} This step identifies the main components of the optimization problem from the input description, organizes them into a structured NL representation, and extracts the metadata needed to build a mathematical formulation. Generally speaking, an optimization problem has four components~\cite{Bertsimas24,williams2013model,Nace20}: (1)~\emph{objectives}, which specify the quantities to be minimized or maximized.  
(2)~\emph{parameters}, which are known problem-specific constants.  
(3)~\emph{variables}, which represent decision points whose optimal values are determined by solving the problem.  
(4)~\emph{constraints}, which are expressed as algebraic relations between variables and parameters, restricting permissible variable values.
 For example, in the problem description of  Figure~\ref{fig:ProductionProblem}, the objectives -- maximizing revenue from product sales and minimizing inventory costs -- are highlighted in \textcolor{magenta}{pink}. The parameters, in \textcolor{darkgreen}{green}, include product and resource types, resource requirements, current inventories, purchase costs, budget, inventory costs, and selling prices. The decision variables, in \textcolor{red}{red}, are the production quantities and additional resource purchases. Finally, the constraints, in \textcolor{darkbrown}{brown}, limit resource use to available and purchased units and restrict purchases to the budget.

Using prompts with few-shot examples, Step~1 extracts the components from the input problem description and organizes them into a structured format. As part of this process, it derives metadata for each parameter and variable, including a symbolic name, a brief description, and its dimension (e.g., scalar, one-dimensional, two-dimensional). Step~1 then substitutes occurrences of the parameters and variables in the original description with markup references to their assigned symbols. For example, references to product and resource types in Figure~\ref{fig:ProductionProblem} become \textsf{\textbackslash param\{TypeProducts\}}  and \textsf{\textbackslash param\{TypeResources\}}, linked to the corresponding metadata. In summary, Step 1 produces a structured NL description that includes: (i)~the extracted objectives, parameters, variables, and constraints, (ii) metadata for each parameter and variable, and (iii)~a rewritten problem description where parameters and variables are replaced with their assigned symbols. A worked example -- omitted here due to space -- is provided in the online appendix~\cite{optigen-supp-a}.

%%%%%

\textbf{Step~2. Transform Data.} In textbook optimization problems, parameter values are often embedded directly in the problem description. In real-world scenarios, however, this is impractical because the required data is typically extensive. Solving such problems therefore requires linking the problem description to external data sources, such as databases. \approach\ enables users to automatically transform tabular data into parameter values. To support this, Step~2 uses the parameter metadata from Step 1 and prompts the user to link a value set for each parameter. This step varies by environment, as it depends on the structure and API of the underlying database. The output of Step 2 is a data file combining parameter metadata with their values. This data is exported from the underlying database and passed to Step 4. If the target specification language is AMPL, the data file is generated in an AMPL-compatible format; for Python, a JSON-formatted file is produced. For example, Figure \ref{fig:ampl-0}(b) shows the solver-ready AMPL data file for the problem in Figure \ref{fig:ProductionProblem}.

%%%%%%%%

\textbf{Step~3. Generate Formal Specification.} In Step~3, the structured  description resulting from Step~1 is transformed into a formal optimization specification. As shown in Figure~\ref{fig:approach}, Step~3 can be invoked in two ways: after Step~1 to create the initial specification, or after Step~4 to refine an existing specification based on solver feedback. Specifically:
\emph{(i)~Initial specification.} When executed for the first time, Step~3 generates a specification using an \llm. The prompt consists of target-language-specific instructions (language syntax rules in the case of AMPL and solver-API guidelines in the case of Python), a set of few-shot examples, and the structured description from Step~1.
\emph{(ii)~Specification refinement.} If Step~4 (solver execution) fails due to compilation or runtime errors, Step~3 is invoked again. Here, \llm\ receives a prompt that extends the original one (i.e., language-specific instructions, few-shot examples, and output from Step~1) with the most recently generated specification and solver feedback. The prompt directs \llm\ to analyze the errors, identify problematic parts, and generate a refined specification.

\textbf{Step~4. Solve the optimization problem.} 
This step uses an optimization solver such as Gurobi~\cite{gurobi} or CPLEX~\cite{cplex}, the formal specification from Step~3, and the solver-ready data from Step~2 to compute a solution. As shown in Figure~\ref{fig:approach}, if attempting to solve the problem results in compilation or runtime errors, \approach\ initiates a refinement loop by sending the current specification and solver feedback to Step~3 (as previously explained) to regenerate the specification. This loop continues until the optimization problem is solved without errors or a predefined iteration limit is reached.
\vspace*{-.15cm}
\section{Empirical Evaluation}
\label{sec:RQ}

We address five research questions, RQ1--RQ5, as presented below. Throughout this section, \emph{structuring} refers to whether Step~1 in Figure~\ref{fig:approach} is applied to the NL problem description, while the \emph{refinement loop} denotes the feedback cycle from Step~4 back to Step~3 in Figure~\ref{fig:approach}. Depending on the chosen formalization language, \emph{specification} refers either to an AMPL model or to a Python program.

\textbf{RQ1 (AMPL vs. Python).} \emph{How do LLM-generated optimization specifications in AMPL and Python compare in terms of  executability  and correctness?}

\textbf{RQ2 (Impact of Structuring).} 
\emph{How does structuring problem descriptions, compared to leaving them unstructured, affect the executability and correctness of LLM-generated optimization specifications?} RQ2 investigates the effect of Step~1 in \approach.

\textbf{RQ3 (Impact of the Refinement Loop).} \emph{How does including the refinement loop vs. excluding it affect  the executability and correctness of LLM-generated optimization specifications?}
RQ3 investigates the effect of the feedback mechanism between Steps~4~and~3 in \approach.

\textbf{RQ4 (Reasoning vs. Instruction-following LLMs).} 
\emph{How does using reasoning vs. instruction-following LLMs affect the executability and correctness of LLM-generated optimization specifications?}

\textbf{RQ5 (Impact of Data Transformation Step).}  \emph{How does including the transform-data step in \approach\ affect the executability and correctness of LLM-generated optimization specifications?}

\subsection{Variants of \approach}
\label{subsec:variants}
We develop eight variants of \approach, listed in Table~\ref{tab:ablation}. Each variant is defined by three binary choices, described as follows: 

\textbf{(i)~The target specification language.}  
As shown in Table~\ref{tab:ablation}, \approach\ has four variants that generate AMPL models and four that generate Python code. We denote the four AMPL variants as \textsc{Ampl1}, \textsc{Ampl2}, \textsc{Ampl3}, and \textsc{Ampl4}, and the four Python variants as \textsc{Python1}, \textsc{Python2}, \textsc{Python3}, and \textsc{Python4}. 

\textbf{(ii)~Whether structuring is applied to the input problem description.}  
Four variants of \approach, marked \emph{structured} in Table~\ref{tab:ablation}, use Step~1 (Figure~\ref{fig:approach}) to structure the input NL description before applying Step~3. The other four variants, marked \emph{unstructured} in Table~\ref{tab:ablation}, skip Step~1 and apply Step~3 directly. 

\textbf{(iii)~Whether the refinement loop is applied.}  
Four variants of \approach, marked \emph{refinement} in Table~\ref{tab:ablation}, include a refinement loop that iterates between Steps~4~and~3 (Figure~\ref{fig:approach}) when compilation or runtime errors occur. The other four variants, marked \emph{one-off} in Table~\ref{tab:ablation}, omit this loop and perform Steps~3 and~4 only once.

\begin{table}[t]
\centering
\caption{Variants of \approach\ defined by three attributes: specification language (AMPL or Python), NL description structuring, and inclusion of the refinement loop.}
\vspace*{-.2cm}
\Description{Definition of eight approach variants, Ampl1 to Ampl4 and Python1 to Python4. For each variant, the table lists the specification language and whether natural-language structuring and the refinement loop are used.}
\label{tab:ablation}
\scriptsize
\renewcommand{\arraystretch}{0.9}
\setlength{\tabcolsep}{2pt}
\scalebox{1.3}{\begin{tabular}{|c|c|l|l|}
\hline
\textbf{Variant Label} & \textbf{Spec. Lang.} & \textbf{\hspace*{1em}Structuring?} & \textbf{Refinement Loop?} \\
\Xhline{1pt}
\textsc{Ampl1} & \multirow{4}{*}{AMPL}   & \xmark\quad  (unstructured)\ & \xmark\quad  (one-off)\\
\textsc{Ampl2} &                          & \xmark\quad  (unstructured)\  & \cmark\quad  (refinement) \\
\textsc{Ampl3} &                          & \cmark\quad  (structured)\    & \xmark\quad  (one-off)\\
\textsc{Ampl4} &                          & \cmark\quad  (structured)\    & \cmark\quad  (refinement)\\
\Xhline{1pt}
\textsc{Python1} & \multirow{4}{*}{Python} & \xmark\quad  (unstructured)\  & \xmark\quad  (one-off)\\
\textsc{Python2} &                          & \xmark\quad  (unstructured)\  & \cmark\quad  (refinement) \\
\textsc{Python3} &                          & \cmark\quad (structured)\    & \xmark\quad  (one-off)\\
\textsc{Python4} &                          & \cmark\quad  (structured)\    & \cmark\quad  (refinement) \\
\hline
\end{tabular}}
%\vspace*{-0.5em}
\end{table}

\subsection{Baseline}
\label{sec:baseline}
To our knowledge, no prior work has addressed generating specifications for AMPL or other optimization DSLs. Existing work on optimization-specification generation mainly uses LLMs to derive Python programs from NL descriptions~\cite{AhmadiTeshniziG24}. Similar to \approach, this line of work applies a refinement loop when the generated specification fails to execute and further includes a step for structuring NL problem descriptions. However, because existing approaches are designed to process NL with embedded data values -- as is common in academic exemplars -- they lack a dedicated data-handling step and do not make parameter and variable metadata explicit. Having a step akin to Step~2 in \approach\ is nonetheless important for producing solver-ready data in large-scale industrial optimization. Motivated by this observation, our baseline adopts a variant of \approach\ that omits an explicit Step~2: data is extracted from the NL descriptions and inlined into the generated Python code.

\subsection{Datasets}
 \label{sec:datasets} 
We use two datasets: (1)~a public dataset, \textsc{Public}, containing 60 NL descriptions of optimization problems across five domains: facility location, network flow, scheduling, portfolio management, and energy optimization. These problem descriptions are drawn from three established sources in the optimization literature~\cite{Bertsimas24,williams2013model,Nace20}. Each problem has a data file and a ground-truth solution. The solver code and formal formulations for these problem descriptions are not publicly available, reducing the likelihood that LLMs encountered directly analyzable versions of these problems during training. (2)~a proprietary dataset, \textsc{Industry}, from Kinaxis, with NL descriptions of six real-world supply-chain planning problems. Each problem includes a data file and a subject-matter-expert solution. Due to confidentiality, the \textsc{Industry} dataset cannot be released, and we are confident it was not used for LLM training. Table~\ref{tab:dataset-stats} provides summary statistics for the two datasets. The problem descriptions in \textsc{Public} include, on average, 1,018.12 characters and 171.61 tokens, while those in \textsc{Industry} contain, on average, 1,729 characters and 354 tokens. Thus, on average, the descriptions in \textsc{Industry} contain 1.7\,$\times$ more characters and 2.06\,$\times$ more tokens than those in \textsc{Public}.

\begin{table}[t]
\centering
\caption{Summary statistics for the \textsc{Public} and \textsc{Industry} datasets used in our evaluation.}
\Description{Summary of the Public and Industry datasets. The table reports the number of optimization problems and the average description length in characters and tokens for each dataset.}
\label{tab:dataset-stats}
\vspace*{-1em}
\scalebox{0.8}{\begin{tabular}{@{}lcc@{}}
\toprule
~ & \textbf{\textsc{Public}} & \textbf{\textsc{Industry}} \\
\midrule
Number of optimization problems in the dataset & 60 & 6 \\
Average number of characters in the problem descriptions  & 1018.12  & 1729 \\
Average number of tokens in the problem descriptions & 171.61 & 354 \\
\bottomrule
\end{tabular}}
% \vspace*{-1.01em}
\end{table}

The \textsc{Public} and \textsc{Industry} datasets consist exclusively of single-objective optimization problems. This characteristic reflects both theoretical and practical considerations: multi-objective problems do not yield a single solution but rather a set of trade-offs (the Pareto set~\cite{metaheuristicsbook}), which requires additional criteria to reduce the ground truth to one solution. Widely used solvers such as Gurobi~\cite{gurobi} and CPLEX~\cite{cplex} reinforce this property, as they do not enumerate the full Pareto frontier but instead return a single Pareto-optimal solution based on user-specified schemes such as weighted-sum aggregation or lexicographic prioritization. For this reason, optimization problems are typically formulated as single-objective ones, and cases that originally involve multiple objectives are reformulated with explicit prioritization to ensure unique ground-truth solutions, making results reproducible and deterministic. That said, \approach\ is agnostic to this factor and supports both multi-objective and single-objective formulations; for example, it can generate  specifications for the illustrative multi-objective problem in Figure~\ref{fig:ProductionProblem}.

\subsection{Evaluation Metrics}
\label{subsec:metric}
We assess both the executability and the correctness of the  specifications generated by \approach. Executability is measured by counting the number of specifications that compile and run without errors, while correctness is measured on a per-specification basis by comparing the solution produced by the successfully executed specification against the ground truth. The specific metrics used to evaluate executability and correctness are as follows:

\textbf{\emph{Execution Success Rate}}: For executability, we report the absolute and relative number of specifications that compile, run without errors, and return a solution. We further report the number of specifications with compilation errors (e.g., syntax errors, invalid declarations) and runtime errors (e.g., infeasibility, unboundedness, unexpected termination).

\textbf{\emph{Relative Error}}:
For correctness, we consider only specifications that compile and produce a solution. For these specifications, we compute the relative error with respect to the numeric, single-objective ground truth. Let $s \in \mathbb{R}$ denote the solution for a given specification, and let $g \in \mathbb{R}$ denote the associated ground truth. The relative error (RelErr) is defined as
$\mathrm{RelErr}(s,g) = \frac{|s - g|}{|g|}$.
RelErr quantifies the deviation of a computed solution from the ground truth, rather than relying on a binary correct/incorrect verdict.

\subsection{Implementation}
Our implementation supports all the \approach\ variants  in Table~\ref{tab:ablation}. It is written in Python~3.10 and built on top of LangChain (v0.2.8), which facilitates structured interactions with LLMs through its expression language (LCEL). For the AMPL variants, we use prompt templates that generate AMPL model (\texttt{.mod}) and data (\texttt{.dat}) files. For the Python variants, the templates produce code targeting the Gurobi optimization API, with input data captured in JSON format. Our  complete implementation is  available online~\cite{neayoughi2025optigen}.

\subsection{Experimental Procedure}
We applied the eight \approach\ variants from Table~\ref{tab:ablation} to the \textsc{Public} and \textsc{Industry} datasets using four LLMs: \textit{GPT-4o}~\cite{openai2024gpt4ocard}, \textit{(GPT) o4-mini}~\cite{openai2025introducing}, \textit{Gemini 1.5-Flash}~\cite{team2024gemini}, and \textit{Gemini 2.5-Pro}~\cite{comanici2025gemini}.All four LLMs were run with their default inference-time hyperparameters. Our GPT experiments were conducted through OpenAI's API, and our Gemini experiments through Google Vertex AI, both under Kinaxis' business subscriptions. GPT-4o and Gemini 1.5-Flash are instruction-following LLMs, while o4-mini and Gemini 2.5-Pro are reasoning LLMs. Considering both types enables us to address RQ4, which examines how instruction-following and reasoning LLMs differ in their ability to generate optimization specifications.

To mitigate random variation, each experiment was repeated \emph{five times}. In total, we evaluated 66 optimization problems with eight \approach\ variants across four LLMs, repeating each experiment five times. This resulted in $66 \times 8 \times 4 \times 5 = \text{10,560}$ formal specification instances. The generation of these instances took \hbox{484 hours ($\approx$20 days).} To solve all specification instances, whether in AMPL or Python, we used the Gurobi solver~\cite{gurobi}.

For the four \approach\ variants with a refinement loop (between Steps~4 and~3 in Figure~\ref{fig:approach}), we capped the number of refinement iterations at five. If errors persisted after five iterations, the specification was recorded as an executability error. Otherwise, refined specifications that compiled and executed without errors within this limit were evaluated using the relative-error metric (Section~\ref{subsec:metric}).

For the baseline (Section~\ref{sec:baseline}), we applied it only to the \textsc{Public} dataset, where embedding parameter values into the NL descriptions of the optimization problems is feasible. Since the baseline does not include an independent data-handling step, it cannot be applied to the  problems in the \textsc{Industry} dataset, which involve large data volumes. Following the  setup in the \approach\ experiments, we used the same four LLMs, repeated the baseline experiments five times, and limited the refinement loop to five iterations. This results in an additional $60 \times 4 \times 5 = \text{1,200}$ specification instances from the baseline for our evaluation.

\subsection{Results}
Table~\ref{tab:results} presents the executability results: the number of successfully executed specifications (\#Exec), the execution success rate (Success), the number of specifications with compilation errors (\#CE), and the number of specifications with runtime errors (\#RE). Here, Success is defined as the percentage of \#Exec over the total number of specifications for which compilation and execution were attempted. We report outcomes across all eight \approach\ variants (Table~\ref{tab:ablation}) and the four LLMs considered (Gemini 1.5-Flash, GPT-4o, Gemini 2.5-Pro, and o4-mini). Each row of Table~\ref{tab:results}(a) is based on 300 specification instances generated from the \textsc{Public} dataset ($60 \text{ problems}\,\times\,5 \text{ runs}$), while each row of Table~\ref{tab:results}(b) is based on 30 instances generated from the \textsc{Industry} dataset ($6 \text{ problems}\,\times\,5 \text{ runs}$).

\begin{table*}[t]
 \centering
 \caption{Executability results for the  specification instances generated by the \approach\ variants on the (a)~\textsc{Public} and (b)~\textsc{Industry} datasets using four LLMs. Metrics: successfully executed instances (\#Exec), success rate (Success), instances with compilation errors (\#CE), and instances with runtime errors (\#RE). Rows labelled ``AMPL vs.\ Python ($\Delta$)'' show the differences between AMPL and Python variants: for \#Exec and Success, \numup{} (resp., \numdown{}) indicates AMPL (resp., Python) superiority; for \#CE and \#RE, negative (resp., positive) values favour AMPL (resp., Python).}
 \Description{Executability results for AMPL and Python variants on the Public and Industry datasets for four LLMs. For each variant, the table reports the number of executable specifications with success rate, the number of compile errors, and the number of runtime errors. Additional rows report AMPL minus Python differences for the same metrics.}
 \label{tab:results}
 \vspace*{-.3cm}
 \scriptsize
 \setlength{\tabcolsep}{2pt}
 % Pastel highlight colours
 \definecolor{amplcol}{HTML}{DDEBF7} % light blue
 \definecolor{pycol}{HTML}{E2F0D9}   % light green
 \noindent\textbf{(a) \textsc{Public} (out of 300 generated instances for each variant and  for each LLM)}
 \begin{minipage}{\textwidth}
\centering
\scalebox{0.95}{\begin{tabular}{|c|c|c|ccc|ccc|ccc|ccc|}
\hline
\multicolumn{3}{|c|}{\multirow{2}{*}{\textbf{Variant}}}
& \multicolumn{3}{c|}{\textbf{Gemini 1.5-Flash}} 
& \multicolumn{3}{c|}{\textbf{GPT-4o}} 
& \multicolumn{3}{c|}{\textbf{Gemini 2.5-Pro}} 
& \multicolumn{3}{c|}{\textbf{o4-mini}} \\
\cline{4-15}
\multicolumn{2}{|c}{} &  
& \#Exec (Success \%) & \#CE & \#RE 
& \#Exec (Success \%) & \#CE & \#RE 
& \#Exec (Success \%) & \#CE & \#RE 
& \#Exec (Success \%) & \#CE & \#RE \\
\Xhline{1.2pt}
\multirow{4}{*}{AMPL} & \multirow{2}{*}{Unstructured} & One-off  
& 121 (40\%) & 52 & 127 
& 122 (41\%) & 105 & 73 
& 236 (79\%) & 23 & 41 
& 225 (75\%) & 48 & 27 \\
\cline{3-3}
&  & Refinement  
& 251 (84\%) & 10 & 39 
& 210 (70\%) & 48 & 42 
& 280 (93\%) & 0 & 20 
& 280 (93\%) & 2  & 18 \\
\cline{2-15}
& \multirow{2}{*}{Structured} & One-off     
& 153 (51\%) & 36 & 111 
& 152 (51\%) & 81 & 67 
& 243 (81\%) & 18 & 39 
& 246 (82\%) & 33 & 21 \\
\cline{3-3}
&  & Refinement   
& 260 (87\%) & 8  & 32 
& 268 (89\%) & 4  & 28 
& 284 (95\%) & 2 & 14 
& 284 (95\%) & 0  & 16 \\
\Xhline{1.2pt}
\multirow{4}{*}{Python} & \multirow{2}{*}{Unstructured} & One-off  
& 187 (62\%) & 2  & 111 
& 227 (76\%) & 1  & 72 
& 281 (94\%) & 0  & 19 
& 268 (89\%) & 1  & 31 \\
\cline{3-3}
&  & Refinement  
& 219 (73\%) & 7  & 74 
& 250 (83\%) & 3  & 47 
& 287 (96\%) & 0  & 13 
& 278 (93\%) & 0  & 22 \\
\cline{2-15}
& \multirow{2}{*}{Structured} & One-off    
& 157 (52\%) & 9  & 134 
& 186 (62\%) & 2  & 112 
& 268 (89\%) & 1  & 31 
& 205 (68\%) & 23 & 72 \\
\cline{3-3}
&  & Refinement    
& 186 (62\%) & 14 & 100 
& 203 (68\%) & 35 & 62 
& 267 (89\%) & 0 & 33 
& 212 (71\%) & 0  & 88 \\
\Xhline{1.2pt}
\multirow{4}{*}{AMPL vs Python ($\Delta$)} 
& \multirow{2}{*}{Unstructured} & One-off    
& \numdown{66 (\%22)} & 50 & 16
& \numdown{105 (\%35)}& 104 & 1 
& \numdown{45 (\%15)} & 23 & 22
& \numdown{43 (\%14)}  & 47  & -4 \\
\cline{3-3}
&                                   & Refinement   
& \numup{32 (\%11)}   & 3  & -35
& \numdown{40 (\%13)}  & 45 & -5
& \numdown{7 (\%2)} & 0 & 7
& \numup{2 (\%0.7)}    & 2  & -4 \\
\cline{2-15}
& \multirow{2}{*}{Structured} & One-off      
& \numdown{4 (\%1)}  & 27 & -23
& \numdown{34 (\%11)} & 79 & -45
& \numdown{25 (\%8)} & 17 & 8
& \numup{41 (\%14)}    & 10 & -51 \\
\cline{3-3}
&                                   & Refinement    
& \numup{74 (\%25)}    & -6 & -68
& \numup{65 (\%22)}    & -31 & -34
& \numup{17 (\%6)} & 2 & -19
& \numup{72 (\%24)}    & 0           & -72 \\
\hline
\end{tabular}}
\end{minipage}

\medskip

% ---------- (b) \textsc{Industry} ----------
\noindent\textbf{(b) \textsc{Industry} (out of 30 generated instances for each variant and for each LLM)}
\begin{minipage}{\textwidth}
\centering
\scalebox{0.95}{\begin{tabular}{|c|c|c|ccc|ccc|ccc|ccc|}\hline
\multicolumn{3}{|c|}{\multirow{2}{*}{\textbf{Variant}}}
& \multicolumn{3}{c|}{\textbf{Gemini 1.5-Flash}} 
& \multicolumn{3}{c|}{\textbf{GPT-4o}} 
& \multicolumn{3}{c|}{\textbf{Gemini 2.5-Pro}} 
& \multicolumn{3}{c|}{\textbf{o4-mini}} \\
\cline{4-15}
\multicolumn{2}{|c}{} & 
& \#Exec (Success \%) & \#CE & \#RE 
& \#Exec (Success \%) & \#CE & \#RE 
& \#Exec (Success \%) & \#CE & \#RE 
& \#Exec (Success \%) & \#CE & \#RE \\
\Xhline{1.2pt}
\multirow{4}{*}{AMPL} & \multirow{2}{*}{Unstructured} & One-off  
& 12 (40\%)  & 8  & 10 
& 12 (40\%)  & 6  & 12  
& 19 (63\%) & 3 & 8 
& 19 (63\%)  & 5  & 6  \\
\cline{3-3}
&  & Refinement  
& 18 (60\%)  & 4  & 8  
& 19 (63\%)  & 6  & 5  
& 23 (77\%) & 1 & 6 
& 24 (80\%)  & 0  & 6  \\
\cline{2-15}
& \multirow{2}{*}{Structured} & One-off     
& 12 (40\%)  & 8  & 10 
& 7 (23\%)   & 14 & 9  
& 14 (47\%) & 8 & 8 
& 19 (63\%)  & 5  & 6  \\
\cline{3-3}
&  & Refinement   
& 24 (80\%)  & 1  & 5  
& 19 (63\%)  & 3  & 8  
& 24 (80\%) & 1 & 5 
& 24 (80\%)  & 0  & 6  \\
\Xhline{1.2pt}
\multirow{4}{*}{Python} & \multirow{2}{*}{Unstructured} & One-off  
& 21 (70\%)  & 0  & 9  
& 12 (40\%)  & 0  & 18 
& 25 (83\%) & 0 & 5 
& 23 (77\%)  & 0  & 7  \\
\cline{3-3}
&  & Refinement  
& 22 (73\%)  & 0  & 8  
& 18 (60\%)  & 0  & 12 
& 25 (83\%) & 0 & 5 
& 23 (77\%)  & 0  & 7  \\
\cline{2-15}
& \multirow{2}{*}{Structured} & One-off    
& 22 (73\%)  & 0  & 8  
& 10 (33\%)  & 1  & 19 
& 23 (77\%) & 0 & 7 
& 21 (70\%)  & 0  & 9  \\
\cline{3-3}
&  & Refinement    
& 24 (80\%)  & 0  & 6  
& 21 (70\%)  & 0  & 9  
& 25 (83\%) & 0 & 5 
& 23 (77\%)  & 0  & 7  \\
\Xhline{1.2pt}
\multirow{4}{*}{AMPL vs Python ($\Delta$)} 
& \multirow{2}{*}{Unstructured} & One-off    
& \numdown{9 (\%30)} & 8   & 1
& 0                 & 6  & -6
& \numdown{6 (\%20)} & 3 & 3
& \numdown{4 (\%13)} & 5   & -1 \\
\cline{3-3}
&                                   & Refinement   
& \numdown{4 (\%13)}  & 4  & 0
& \numup{1 (\%3)}  & 6  & -7
& \numdown{2 (\%7)} & 1 & 1
& \numup{1 (\%3)}  & 0  & -1 \\
\cline{2-15}
& \multirow{2}{*}{Structured} & One-off      
& \numdown{10 (\%33)} & 8  & 2
& \numdown{3 (\%10)}  & 13 & -10
& \numdown{9 (\%30)} & 8 & 1
& \numdown{2 (\%7)}& 5  & -3 \\
\cline{3-3}
&                                   & Refinement    
& 0                    & 1           & -1
& \numdown{2 (\%7)}  & 3           & -1
& \numdown{1 (\%3)} & 1 & 0
& \numup{1 (\%3)}     & 0           & -1 \\
\hline
\end{tabular}}
\end{minipage}
\end{table*}

\begin{table*}[t]
\centering
\caption{Correctness results for the optimization specifications generated by the \approach\ variants on (a) \textsc{Public} and (b) \textsc{Industry} datasets using four LLMs. Metrics: Mean, Median (Med), Std of relative error, and number of zero-error solutions (\#Zero).} 
\Description{Correctness results for AMPL and Python variants on the Public and Industry datasets for four LLMs. For each variant and LLM, the table reports the mean, median, and standard deviation of relative error over executable specifications, plus the count of zero-error solutions. Additional rows report AMPL minus Python differences for these statistics.}
\label{tab:relerror-stats}
\vspace*{-.3cm}
\scriptsize
\renewcommand{\arraystretch}{1.05}
\setlength{\tabcolsep}{2pt}

% Pastel highlight colours

% ---------- (a) \textsc{Public} ----------
\noindent\textbf{(a) \textsc{Public}}
\begin{minipage}{\textwidth}
\centering
\scalebox{1.05}{\begin{tabular}{|c|c|c|cccc|cccc|cccc|cccc|}
\hline
\multicolumn{3}{|c|}{\multirow{2}{*}{\textbf{Variant}}}
& \multicolumn{4}{c|}{\textbf{Gemini 1.5-Flash}}
& \multicolumn{4}{c|}{\textbf{GPT-4o}}
& \multicolumn{4}{c|}{\textbf{Gemini 2.5-Pro}}
& \multicolumn{4}{c|}{\textbf{o4-mini}} \\
\cline{4-19}
\multicolumn{2}{|c}{} &
& Mean & Med & Std & \#Zero
& Mean & Med & Std & \#Zero
& Mean & Med & Std & \#Zero
& Mean & Med & Std & \#Zero \\
\Xhline{1.2pt}
\multirow{4}{*}{AMPL} & \multirow{2}{*}{Unstructured} & One-off
& 0.59 & 0 & 2.05 & 84
& 0.59 & 0 & 1.98 & 82
& 0.48 & 0 & 2.84 & 176
& 0.70 & 0 & 2.70 & 161 \\
\cline{3-3}
& & Refinement
& 1.40 & 0.03 & 4.87 & 106
& 0.88 & 0 & 2.91 & 134
& 0.59 & 0 & 2.76 & 165
& 0.76 & 0 & 3.08 & 197 \\
\cline{2-19}
& \multirow{2}{*}{Structured} & One-off
& 1.79 & 0 & 6.27 & 79
& 1.20 & 0 & 3.56 & 78
& 0.93 & 0 & 4.88 & 143
& 0.86 & 0 & 3.71 & 153 \\
\cline{3-3}
& & Refinement
& 0.74 & 0 & 2.97 & 179
& 1.27 & 0 & 4.33 & 127
& 0.10 & 0 & 0.25 & 203
& 0.86 & 0 & 3.76 & 166 \\
\Xhline{1.2pt}
\multirow{4}{*}{Python} & \multirow{2}{*}{Unstructured} & One-off
& 0.70 & 0 & 3.03 & 96
& 3.24 & 0 & 15.74 & 127
& 0.75 & 0 & 3.07 & 202
& 0.80 & 0 & 3.15 & 148 \\
\cline{3-3}
& & Refinement
& 0.81 & 0 & 2.91 & 109
& 2.72 & 0 & 12.80 & 133
& 0.73 & 0 & 3.04 & 209
& 0.78 & 0 & 3.09 & 183 \\
\cline{2-19}
& \multirow{2}{*}{Structured} & One-off
& 1.58 & 0 & 7.33 & 87
& 7.10 & 0 & 86.15 & 131
& 1.82 & 0 & 9.08 & 154
& 1.27 & 0 & 6.71 & 140 \\
\cline{3-3}
& & Refinement
& 1.34 & 0 & 6.74 & 110
& 1.01 & 0 & 3.26 & 122
& 0.74 & 0 & 2.90 & 162
& 0.25 & 0 & 1.34 & 167 \\
\hline
\end{tabular}}
\end{minipage}

\medskip

% ---------- (b) \textsc{Industry} ----------
\noindent\textbf{(b) \textsc{Industry}}
\begin{minipage}{\textwidth}
\centering
\scalebox{1.05}{\begin{tabular}{|c|c|c|cccc|cccc|cccc|cccc|}
\hline
\multicolumn{3}{|c|}{\multirow{2}{*}{\textbf{Variant}}}
& \multicolumn{4}{c|}{\textbf{Gemini 1.5-Flash}}
& \multicolumn{4}{c|}{\textbf{GPT-4o}}
& \multicolumn{4}{c|}{\textbf{Gemini 2.5-Pro}}
& \multicolumn{4}{c|}{\textbf{o4-mini}} \\
\cline{4-19}
\multicolumn{2}{|c}{} &
& Mean & Med & Std & \#Zero
& Mean & Med & Std & \#Zero
& Mean & Med & Std & \#Zero
& Mean & Med & Std & \#Zero \\
\Xhline{1.2pt}
\multirow{4}{*}{AMPL} & \multirow{2}{*}{Unstructured} & One-off
& 0.07 & 0 & 0.23 & 11
& 0.67 & 0 & 1.48 & 8
& 0.16 & 0 & 0.33 & 15
& 0.18 & 0 & 0.41 & 15 \\
\cline{3-3}
& & Refinement
& 0.15 & 0 & 0.29 & 14
& 0.18 & 0 & 0.34 & 12
& 0.35 & 0 & 0.86 & 16
& 0.15 & 0 & 0.37 & 20 \\
\cline{2-19}
& \multirow{2}{*}{Structured} & One-off
& 0.07 & 0 & 0.25 & 11
& 0.14 & 0 & 0.38 & 5
& 0.00 & 0 & 0.00 & 14
& 0.04 & 0 & 0.16 & 18 \\
\cline{3-3}
& & Refinement
& 0.22 & 0 & 0.41 & 18
& 0.19 & 0 & 0.82 & 17
& 0.47 & 0 & 1.13 & 18
& 0.12 & 0 & 0.27 & 20 \\
\cline{2-19}
\Xhline{1.2pt}
\multirow{4}{*}{Python} & \multirow{2}{*}{Unstructured} & One-off
& 0.14 & 0 & 0.31 & 17
& 1.78 & 0 & 5.05 & 10
& 0.15 & 0 & 0.30 & 20
& 0.17 & 0 & 0.46 & 20 \\
\cline{3-3}
& & Refinement
& 0.18 & 0 & 0.34 & 17
& 0.50 & 0 & 1.22 & 10
& 0.20 & 0 & 0.35 & 18
& 0.20 & 0 & 0.47 & 19 \\
\cline{2-19}
& \multirow{2}{*}{Structured} & One-off
& 0.14 & 0 & 0.30 & 18
& 1.87 & 0 & 3.70 & 14
& 0.14 & 0 & 0.31 & 19
& 0.08 & 0 & 0.27 & 19 \\
\cline{3-3}
& & Refinement
& 0.19 & 0 & 0.33 & 18
& 3.08 & 0.01 & 6.93 & 10
& 0.16 & 0 & 0.31 & 19
& 0.26 & 0 & 0.82 & 20 \\
\hline
\end{tabular}}
\end{minipage}
\end{table*}

The last four rows of Tables~\ref{tab:results}(a) and~\ref{tab:results}(b), marked ``AMPL vs.~Python ($\Delta$)'', show the differences between each AMPL variant and its Python counterpart. For \#Exec and Success, \numup{} means the AMPL variant achieves a higher execution success rate, while \numdown{} means the opposite. For \#CE and \#RE, negative values mean the AMPL variant yields fewer errors, while positive values mean the opposite.

Table~\ref{tab:relerror-stats} presents the correctness results, including the mean, median (Med), and standard deviation (Std) of the relative-error metric defined in Section~\ref{subsec:metric}, as well as the number of specification instances yielding a perfect solution, i.e., zero relative error (\#Zero).

To address our research questions using the results in Tables~\ref{tab:results} and~\ref{tab:relerror-stats}, we apply the following statistical tests: $Z$-test~\cite{montgomery2019applied}, Mann-Whitney $U$ test~\cite{Mann-Whitney}, and Vargha-Delaney effect size $\hat{A}_{12}$~\cite{vargha2000critique}. Specifically, we use the $Z$-test for our executability metric, execution success rate (Success).  We report $Z$- and $p$-values at the 5\% significance level, concluding that variant \emph{A} (the first variant in the comparison) outperforms \emph{B} (the second variant) if $Z > 0$ and $p < 0.05$. Otherwise, variant \emph{B} outperforms  \emph{A} if $Z < 0$ and $p < 0.05$. 

To compare variants on the relative-error metric, we use the $U$~test together with $\hat{A}_{12}$. Relative error is computed only for successfully executed specifications, so the sample size for each variant depends on its execution success rate. As a non-parametric method, the $U$~test is appropriate for comparing two distributions of unequal sizes. Comparisons are performed at the 5\% significance level. Since smaller errors indicate better accuracy, we conclude that \emph{A} (the first variant in the comparison) outperforms \emph{B} (the second variant) if $\hat{A}_{12} < 0.5$, with thresholds of 0.44 (small), 0.36 (medium), and 0.29 (large). The difference is negligible when  $0.44 < \hat{A}_{12} < 0.5$.

Due to space constraints, statistical test results for RQ1--RQ4 are provided online~\cite{optigen-supp-b}, and we summarize them when answering these RQs. Results for RQ5 are reported in the paper.

\vspace*{.5em}\textbf{RQ1 (AMPL vs. Python).}
We statistically compare the four AMPL-based variants of \approach\ with their corresponding Python-based variants using the results in Tables~\ref{tab:results} and~\ref{tab:relerror-stats}.

Across the 32 comparisons (4 variant pairs $\times$ 4 LLMs $\times$ 2 datasets), assessing AMPL-based and Python-based variants in terms of relative error, the Python-based variants never outperform the AMPL-based ones. In contrast, the AMPL-based variants significantly outperform the Python-based variants in five comparisons, although the effect sizes are negligible, small, or medium.

For execution success rate, when structuring is applied to NL descriptions and the refinement loop is included, AMPL-based variants either significantly outperform or perform comparably to the Python-based variants. In contrast, when either structuring or refinement is excluded, the results are mixed: in two comparisons, the AMPL-based variants perform significantly better; in eleven comparisons, the Python-based variants outperform the AMPL-based ones; and in the rest, neither shows a significant advantage.

\begin{tcolorbox}[breakable,colback=gray!10!white,colframe=black!75!black,boxrule=0.5mm,arc=1mm,left=1mm,right=1mm,top=1mm,bottom=1mm,fonttitle=\bfseries]
\textbf{The answer to RQ1} is that deriving optimization specifications in Python yields no statistically significant improvements in correctness compared to AMPL. In contrast, in some comparisons AMPL yields statistically significant improvements in correctness. Regarding executability, AMPL shows significant gains over Python when specifications are derived from structured descriptions and refined iteratively. In other scenarios, however, Python-based variants may outperform AMPL-based ones in terms of executability.
\end{tcolorbox}

\textbf{RQ2 (Impact of Structuring).} To address RQ2, we compare \approach\ variants that include the structuring step (Step~1 in Figure~\ref{fig:approach}) with variants that skip it. 
As shown in Table~\ref{tab:ablation}, this entails comparing \textsc{Ampl1} (unstructured) with \textsc{Ampl3} (structured), \textsc{Ampl2} (unstructured) with \textsc{Ampl4} (structured), \textsc{Python1} (unstructured) with \textsc{Python3} (structured), and \textsc{Python2} (unstructured) with \textsc{Python4} (structured). Similar to RQ1, there are 32 comparison combinations across four LLMs and two datasets. Unlike RQ1, however, these comparisons focus on the effect of structuring. Our results show that, for relative error, variants with structuring significantly outperform variants without structuring in five comparisons, with negligible, small, or medium effect sizes. None of the variants without structuring achieve statistically significant improvement over variants with structuring for relative error.

For execution success rate, AMPL variants with structuring outperform AMPL variants without structuring in four comparisons; in the remaining cases, neither group shows a clear advantage. In contrast, comparisons between Python variants with and without structuring yield mixed results, with each outperforming the other in different scenarios.

\begin{tcolorbox}[breakable,colback=gray!10!white,colframe=black!75!black,boxrule=0.5mm,arc=1mm,left=1mm,right=1mm,top=1mm,bottom=1mm,fonttitle=\bfseries]
\textbf{The answer to RQ2} is that structuring problem descriptions prior to deriving AMPL models never results in a disadvantage compared to variants without structuring in terms of executability and correctness. AMPL variants with structuring either yield statistically significant improvements in these metrics or perform on par with variants without structuring, showing no statistically significant difference. In contrast, for Python variants, structuring the NL descriptions does not provide consistent improvement over variants without structuring.
\end{tcolorbox}

\textbf{RQ3 (Impact of the Refinement Loop).} To address RQ3, we compare \approach\ variants that include the refinement loop between Step~4 and Step~3 in Figure~\ref{fig:approach} with variants that generate specifications without it. As shown in Table~\ref{tab:ablation}, this entails comparing \textsc{Ampl1} (one-off) with \textsc{Ampl2} (refinement), \textsc{Ampl3} (one-off) with \textsc{Ampl4} (refinement), \textsc{Python1} (one-off) with \textsc{Python2} (refinement), and \textsc{Python3} (one-off) with \textsc{Python4} (refinement). Similar to RQ1 and RQ2, there are 32 comparison combinations. Unlike RQ1 and RQ2, however, these comparisons focus on the effect of the refinement loop. Our results show that variants with the refinement loop significantly outperform variants without it in 14 cases for execution success rate and three cases for relative error. The effect sizes for all statistically significant differences are negligible or small. None of the variants without the refinement loop show statistically significant improvement over variants with the refinement loop across executability and correctness metrics.

\begin{tcolorbox}[breakable,colback=gray!10!white,colframe=black!75!black,boxrule=0.5mm,arc=1mm,left=1mm,right=1mm,top=1mm,bottom=1mm,fonttitle=\bfseries]
\textbf{The answer to RQ3} is that variants with the refinement loop never result in a disadvantage compared to variants without the refinement loop. Variants with the refinement loop either yield statistically significant improvements in executability and correctness metrics or perform on par with variants without the refinement loop, showing no statistically significant difference.
\end{tcolorbox}

\textbf{RQ4 (Reasoning vs. Instruction-following LLMs).} To address RQ4, we 
evaluate each variant of \approach\ in two settings: first with two reasoning LLMs (Gemini 2.5-Pro and o4-mini) and then with two instruction-following LLMs (Gemini 1.5-Flash and GPT-4o). For each variant, we contrast the average results obtained with reasoning LLMs against the average results of the same variant with instruction-following LLMs. Across the 16 comparisons of the eight variants over our two datasets, reasoning LLMs significantly outperform instruction-following LLMs in 13 cases for execution success rate and eight cases for relative error. The effect sizes for all statistically significant differences are negligible, small, or medium. No significant improvement is observed for instruction-following LLMs over reasoning LLMs.

\begin{tcolorbox}[breakable,colback=gray!10!white,colframe=black!75!black,boxrule=0.5mm,arc=1mm,left=1mm,right=1mm,top=1mm,bottom=1mm,fonttitle=\bfseries]
\textbf{The answer to RQ4} is that reasoning LLMs never result in a disadvantage compared to instruction-following LLMs when generating optimization specifications. Reasoning LLMs either yield statistically significant improvements in executability and correctness or perform on par with instruction-following LLMs, showing no statistically significant difference.
\end{tcolorbox}

\begin{figure}[t]
    \centering
    \Description{Plot that compares Ampl4 and Python4 on the Public and Industry datasets, reporting average execution success rate, average number of zero-error solutions, and average relative error for reasoning LLMs.}
    \hspace*{-.5em}\includegraphics[width=1.05\linewidth]{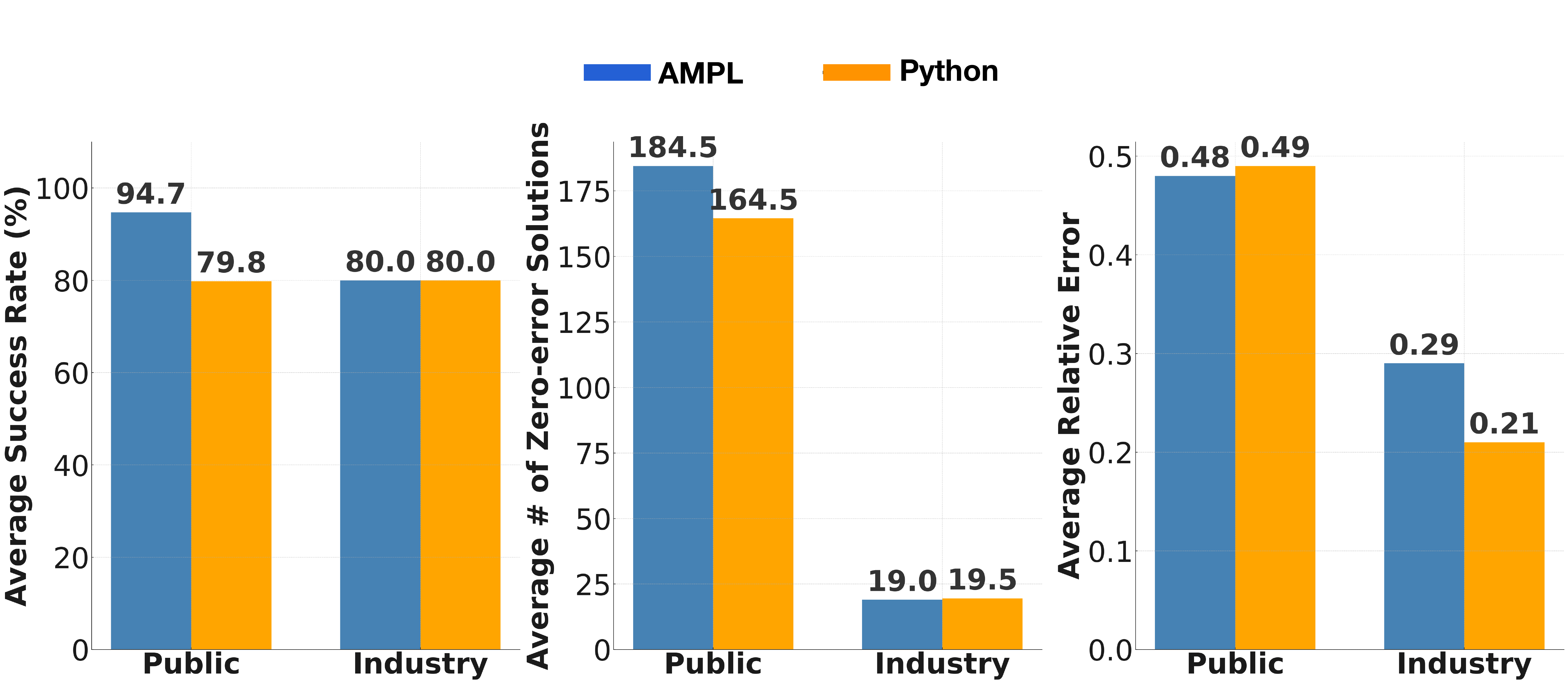}
    \vspace*{-.5cm}
    \caption{Comparison of \approach\ variants that generate AMPL models and Python code from structured descriptions with refinement loops, showing average execution success rate, average number of zero-error solutions, and average relative error when applied with reasoning LLMs on the \textsc{Public} and \textsc{Industry} datasets.}
    \label{fig:results_comparison}
    \vspace*{-.5cm}
\end{figure}

\noindent
\tikz[baseline=(X.base)]{
  \node[
    rounded corners=1.5pt,
    inner sep=2pt,
    fill=blue!15
  ] (X) {\textcolor{red}{{\Large$\blacktriangleright$}}\ \textbf{Main Finding from RQ1--RQ4}.};
} Our results for RQ1 to RQ4 indicate that the most effective variant of \approach\ is \textsc{Ampl4}, as shown in Table~\ref{tab:ablation}, when used with reasoning LLMs. This variant structures the input problem description, then generates and iteratively refines the AMPL model. Figure~\ref{fig:results_comparison} compares the average execution success rate, the average number of zero-error solutions, and the average relative error yielded by the two reasoning LLMs for \textsc{Ampl4}, with those obtained by its Python-based counterpart, \textsc{Python4}, on the \textsc{Public} and \textsc{Industry} datasets.

On the \textsc{Public} dataset, \textsc{Ampl4} achieves a 94.7\% average success rate, and 80\% on \textsc{Industry}. Among the (AMPL) models that execute, an average of 65\% (184.5 out of 284) yield exact solutions for \textsc{Public}, and 79\% (19 out of 24) for \textsc{Industry}. In addition, \textsc{Ampl4} yields an average relative error of 0.48 for \textsc{Public} and 0.29 for \textsc{Industry}. Across all problems in both datasets -- except for a single case in \textsc{Industry} -- \textsc{Ampl4} produces at least one successfully compiled AMPL model within five runs.

Compared to \textsc{Python4}, \textsc{Ampl4} shows advantages in both average execution success rate and the average number of zero-error solutions. On the \textsc{Public} dataset, \textsc{Ampl4} outperforms \textsc{Python4} with a 14.9\% higher success rate (94.7\% vs.\ 79.8\%) and generates on average 20 more exact solutions (184.0 vs.\ 164.5), an improvement of 12\%. On the \textsc{Industry} dataset, \textsc{Ampl4} and \textsc{Python4} perform comparably in terms of executability (80\% each) and correctness (19.0 vs.\ 19.5 exact solutions on average). 
Since the Python and AMPL variants have comparable success rates in the comparisons, relative error computed over successful executions does not favour one over the other.

For average relative error, the result patterns diverge slightly: on the \textsc{Public} dataset, \textsc{Ampl4} matches \textsc{Python4} (0.48 vs.\ 0.49), while on the \textsc{Industry} dataset it shows a slightly higher error (0.29 vs.\ 0.21). Overall, statistical tests indicate that the average reduction in relative error of \textsc{Python4} over \textsc{Ampl4} is not significant (see~\cite{optigen-supp-b}).

\begin{tcolorbox}[breakable,colback=gray!10!white,colframe=black!75!black,boxrule=0.5mm,arc=1mm,left=1mm,right=1mm,top=1mm,bottom=1mm,fonttitle=\bfseries]
\textbf{Takeaway.} On the question of whether LLM-generated models in a DSL like AMPL could pose a disadvantage compared to code in a mainstream language like Python, we find that, in our study context, \textbf{AMPL models generated by reasoning LLMs with structuring and iterative refinement are as reliable as Python code generated similarly, and sometimes better.}
\end{tcolorbox}

\textbf{RQ5 (Impact of Data Transformation Step).} 
We compare the best AMPL and Python variants of \approach, namely \textsc{Ampl4} and \textsc{Python4}, as identified in RQ1–4, with the baseline  discussed in Section~\ref{sec:baseline}. Table~\ref{tab:rq5}(a) reports the average values for the baseline in terms of the \#Exec, Success, RelErr, and \#Zero metrics across the four LLMs considered. Table~\ref{tab:rq5}(b) presents the statistical tests comparing \textsc{Ampl4} and \textsc{Python4} against the baseline.
%%%%%%%%%%%%%%%%%%%%%
%tables
% RQ5 (Impact of Data Transformation Step)

\begin{table}[t]
\centering
\caption{Best AMPL and Python variants of \approach\ vs. the baseline on the \textsc{Public} dataset.
Blue cells indicate significant improvements over the baseline; no significant differences were observed where the baseline outperforms \approach.}
\Description{Two-part table for the Public dataset. Part (a) reports baseline executability and relative error statistics for each LLM, and the number of zero-error cases. Part (b) reports one-sided z-test results for Ampl4 vs baseline and Python4 vs baseline, with p-values, z statistics, and A12 effect sizes for success and relative error.}
\label{tab:rq5}
\scriptsize
\definecolor{amplcol}{HTML}{DDEBF7} % light blue
\setlength{\tabcolsep}{3pt}
\vspace*{-.5em}
% ---------- Table (a) ----------
\noindent\textbf{(a) Executability and correctness results for the baseline on the \textsc{Public} dataset}\\[0.1cm]
\scalebox{1}{
\begin{tabular}{|l|cccc|}
\hline
\textbf{Metric} & \textbf{Gemini 1.5 Flash} & \textbf{GPT-4o} & \textbf{Gemini 2.5 Pro} & \textbf{o4-mini} \\
\Xhline{1.1pt}
\#Exec (Succ.\%) & 171 (57\%) & 206 (69\%) & 276 (92\%) & 275 (92\%) \\
\cline{1-5}
Mean (RelErr) & 1.45 & 2.83 & 0.57 & 0.61 \\
Med (RelErr) & 0 & 0 & 0 & 0 \\
Std (RelErr) & 7.01 & 13.35 & 2.73 & 2.75 \\
\#Zero (RelErr) & 98 & 129 & 201 & 156 \\
\hline
\end{tabular}
} \\[0.15cm]
% ---------- Table (b) ----------
\noindent\textbf{(b) Statistical tests comparing \textsc{Ampl4} and \textsc{Python4} against the baseline; all p-values are rounded to two decimal places}\\[0.1cm]
\scalebox{0.9}{
\begin{tabular}{|l||cc|cc||cc|cc|}
\hline
\multirow{3}{*}{\textbf{LLM}}
  & \multicolumn{4}{c||}{\textsc{Ampl4} vs. Baseline}
  & \multicolumn{4}{c|}{\textsc{Python4} vs. Baseline} \\
\cline{2-9}
  & \multicolumn{2}{c|}{Success} & \multicolumn{2}{c||}{RelErr}
  & \multicolumn{2}{c|}{Success} & \multicolumn{2}{c|}{RelErr} \\
\cline{2-9}
  & \textit{p-val} & Z & \textit{p-val} & $\hat{A}_{12}$
  & \textit{p-val} & Z & \textit{p-val} & $\hat{A}_{12}$ \\
\hline
\textbf{Gemini 1.5 Flash} & \cellcolor{amplcol}\textbf{0.00} & \cellcolor{amplcol}\textbf{4.88}
         & \cellcolor{amplcol}\textbf{0.00} & \cellcolor{amplcol}\textbf{0.37(M)}
         & 0.06 & -1.84 & 0.17 & 0.48 \\
\textbf{GPT-4o}  & \cellcolor{amplcol}\textbf{0.00} & \cellcolor{amplcol}\textbf{6.21}
         & 0.65 & 0.51
         & 0.60 & -0.26 & 0.99 & 0.57 \\
\textbf{Gemini 2.5 Pro} & 0.10 & 1.31
         & 1.00 & 0.59
         & 0.89 & -1.25
         & 0.34 & 0.49 \\
\textbf{o4-mini }     & 0.07 & 1.46
         & 0.21 & 0.48
         & 1.00 & -6.58
         & 1.00 & 0.59 \\
\hline
\end{tabular}
}
\vspace*{-1em}
\end{table}

%%%%%%%%%%%%%%%%%%%%%

Comparing the \approach\ results in Tables~\ref{tab:results} and~\ref{tab:relerror-stats} with the baseline results in Table~\ref{tab:rq5}(a) shows that,  on average, \textsc{Ampl4} increases execution success by 14\%, reduces the average RelErr by 46\%, and improves the number of zero-error cases by 23 compared to the baseline. As shown in Table~\ref{tab:rq5}(b), \textsc{Ampl4} shows a statistically significant improvement over the baseline in execution success for two LLMs, and in relative error for one LLM, with a medium effect size. No statistically significant improvement is observed for the baseline over \textsc{Ampl4}. In contrast, \textsc{Python4} generally performs on par with the baseline, as neither shows a statistically significant improvement over the other.

\begin{tcolorbox}[breakable,colback=gray!10!white,colframe=black!75!black,boxrule=0.5mm,arc=1mm,left=1mm,right=1mm,top=1mm,bottom=1mm,fonttitle=\bfseries]
\textbf{The answer to RQ5} is that the best-performing AMPL variant of \approach\ outperforms the baseline in executability and correctness, with statistically significant gains in executability and correctness. The baseline shows no statistically significant gains over our best Python variant.
\end{tcolorbox}

\subsection{Validity Considerations}\label{sec:threats}
\mbox{}\indent\textbf{Internal validity.} All \approach\ variants use a common implementation stack and solver (Gurobi). Every configuration was repeated five times, with refinement loops, where present, capped at five iterations. The baseline was applicable only to the \textsc{Public} dataset but was matched to \approach\ in solver, loop caps, and repetitions.  

\textbf{Construct validity.} Our metrics focus on executability and objective accuracy, but omit human-centred aspects such as readability and maintainability. Generating DSL models rather than code is a step towards these qualities, but substantiating such benefits would require user studies with dedicated measurement constructs.

\textbf{Conclusion validity.} Executability was analyzed with Z-tests, while relative errors were compared using the Mann-Whitney $U$ test and the Vargha-Delaney $\hat{A}_{12}$ effect size to avoid overstating significance.

\textbf{External validity.} Our evaluation spans 60 benchmark and six industrial problems across diverse scales and domains. We consider both reasoning and instruction-following LLMs, providing broad coverage.
Nevertheless, our study is scoped to mathematical optimization using AMPL and Python. While our results point to interesting dynamics between LLM-generated DSL models and general-purpose code, they may not generalize to other languages or domains. Moreover, our findings reflect specific LLM capabilities at the time of writing; as LLMs evolve, trade-offs between LLM-generated models and code may shift, warranting re-evaluation.

\textbf{Reliability.} We release the \approach\ code, prompts, and the \textsc{Public} dataset~\cite{neayoughi2025optigen}, but cannot share the \textsc{Industry} dataset. This reduces pretraining-leakage risk but limits replication. To improve reliability, we document our experimental procedure in detail.

\section{Lessons Learned}\label{sec:lessons}
We believe that the comparable, and in some cases superior, quality of AMPL models compared to Python code in terms of executability and correctness, as observed in Section~\ref{sec:RQ}, can be explained by two factors. \emph{First}, while LLMs may not have been exposed to as many AMPL models as Python programs, the syntax of AMPL closely matches well-defined characteristics of the mathematical optimization domain. This domain has extensive reference material likely included in LLM pretraining. With proper introduction of AMPL syntax through prompting, we were able to get LLMs to generate highly accurate AMPL models. \textbf{The lesson is that} \emph{although DSLs are less prevalent in LLM training data compared to general-purpose programming languages, when DSLs capture recurring structural patterns characteristic of a well-defined domain, the quality of LLM-generated DSL models can match -- or even surpass -- that of LLM-generated code in broad-based programming languages.}

\emph{Second}, \approach\ aligns with industrial practice by distinguishing data from problem formulation, delegating to LLMs only the task of deriving formal specifications from informal (NL) descriptions, rather than both data organization and specification derivation. Real-world problems usually involve too much data to embed directly in problem descriptions, making combined representations impractical and rare. By separating data from the problem description, LLMs can focus on a clear and well-defined task -- specification derivation -- rather than juggling this task with data organization.

The baseline approach evaluated in RQ5 does not make this distinction: LLMs are tasked with producing both solver-ready data and (Python-based) problem formulations. This results in lower accuracy, compared with the best-performing AMPL variant of \approach. \textbf{The lesson is that} \emph{distinguishing between data and problem description -- a common  practice in industrial settings -- enables LLMs to generate more accurate formal specifications.} 
\section{Related Work}\label{sec:related}
Automated assistance for extracting models from text has been studied prior to LLMs, e.g.,~\cite{Yue11,AroraSBZ16,AroraSNB19}. With LLMs, this research has accelerated, producing more effective methods for generating various models, including goal models~\cite{Chen23}, domain models~\cite{YangCCMV24,silva2024application,Chen25}, sequence diagrams~\cite{FerrariAA24,JahanHGRRRS24}, and activity diagrams~\cite{khamsepour2025impact,Chen25}. 
Wang et al.~\cite{WangW0CSK23} propose grammar prompting to improve few-shot DSL generation by pairing each example with a small grammar fragment that captures the relevant syntax rules. The LLM first predicts such a grammar for the input and then generates the output under this grammar's rules. In contrast, \approach\ does not impose a formal grammar during generation. Instead, it relies on few-shot examples and domain-specific syntax guidelines, and it introduces a structuring step that identifies problem components prior to generating AMPL models. Recently, Joel et al.~\cite{JoelWF25} have conducted a systematic review of LLM-based code generation for DSLs. Their findings identify iterative feedback from external tools, such as compilers or solvers, as a technique that improves LLM accuracy; this aligns with the use of solver diagnostics in our work.

LLMs have also been used to generate logic-based artifacts, such as  temporal-logic formulas~\cite{CoslerHMST23,ChenGZF23}, OCL constraints~\cite{AbukhalafHK24}, and assertion-based postconditions~\cite{EndresFCL24}. These artifacts can be validated only in combination with a host artifact (e.g., a formal model for temporal logics, Java code for postconditions, or a UML model for OCL), making the effectiveness of their automated derivation inseparable from that context and thus precluding direct compiler- or solver-style diagnostics. Closer to our setting, work on generating optimization specifications~\cite{AhmadiTeshniziG24,ZhangWGWLYY24} does yield solver-executable artifacts with solver feedback driving refinement. However, these approaches do not explicate how optimization data is acquired and bound, and couple such tasks with specification extraction, making them ill-suited for industrial-scale datasets. In contrast, informed by practice, \approach\ introduces a dedicated data-handling step, decoupled from specification extraction, to ensure reliable data binding at scale.
 Finally, to our knowledge, no prior work on extracting optimization specifications from text provides a systematic comparison of LLMs in DSLs versus general-purpose languages, as we do, or ablates components to quantify their impact. 

\begin{acks}
We gratefully acknowledge financial support from Mitacs, Kinaxis, and NSERC of Canada (Discovery Program). We thank the UX and AI teams at Kinaxis for constructive discussions and feedback.
\end{acks}

\balance

\bibliographystyle{ACM-Reference-Format}
\bibliography{bibliography}
\end{document}